\documentclass[utf8]{frontiersinFPHY_FAMS} 
\setcitestyle{square} 
\usepackage{url,hyperref,lineno,microtype,subcaption}
\usepackage[onehalfspacing]{setspace}
\usepackage{pifont}
\usepackage{hyperref}
\usepackage{graphicx,graphbox}
\usepackage[utf8]{inputenc}
\usepackage{amsmath}
\usepackage{verbatim}
\usepackage{amssymb}
\usepackage{listings}
\usepackage{hyperref}
\usepackage{mathtools}
\usepackage{ifthen}
\usepackage{color}
\usepackage{xspace}
\usepackage{tikz-cd}
\usepackage{float}
\usepackage{caption}
\usepackage{subcaption}
\usepackage{bm}

\newcommand\taucoup{\tau_\text{coup}}
\newcommand\taumix{\tau_\text{mix}}

\newcommand{\pit}[1]{\pi^{\{#1\}}}

\newcommand{\Tpath}{T_{\text{path}}}
\newcommand{\betapath}{\beta_{\text{path}}}

\newcommand{\Tcoup}{T_{\text{coup}}}
\newcommand{\betacoup}{\beta_{\text{coup}}}

\newcommand{\Pcoup}[2]{P^{\text{coup}}[(#1),(#2)]}    

\newcommand{\etacoup}{\eta_{\text{coup}}}
\newcommand{\etapath}{\eta_{\text{path}}}

\newcommand{\TSpinGlass}{T_{\text{SG}}}

\newcommand{\TGriffiths}{T_{\text{Griffiths}}}

\newcommand{\sigmaa}{\sigmavec^A}
\newcommand{\sigmab}{\sigmavec^B}

\newcommand{\standard}{random-share\xspace}
\newcommand{\EA}{Edwards--Anderson model\xspace}

\newcommand{\subcap}[1]{{\emph{#1}:}}

\newcommand{\REF}[2][]{
	\ifthenelse{\equal {#1} {}}{Ref.~\cite{#2}}{Ref.~\cite[#1]{#2}}}
\newcommand{\FUNCTION}[2][]{\sub{#2} \glb #1 \grb }

\newcommand{\SET}[1]{\{#1\}}

\newcommand{\sub}[1]{\texttt{#1}}
\newcommand{\eq}[1]{eq.~(\ref{#1})}

\newcommand{\fig}[1]{Fig.~\ref{#1}}
\newcommand{\figg}[1]{Figure~\ref{#1}}
\newcommand{\figtwo}[2]{Figs~\ref{#1} and~\ref{#2}}
\newcommand{\quot}[1]{``#1''}
\newcommand{\tab}[1]{Table~\ref{#1}}

\newcommand{\sect}[1]{Sec.~\ref{#1}} 
\newcommand{\subsect}[1]{Subsec.~\ref{#1}}

\newcommand{\SECT}[1]{Section~\ref{#1}}

\newcommand{\NCAL}{\mathcal{N}}  
\newcommand{\OCAL}{\mathcal{O}}  
\newcommand{\expa}[1]{\mathrm{e}^{#1}}   
\newcommand{\expb}[1]{\exp \glb #1 \grb} 
\newcommand{\expc}[1]{\exp \glc #1 \grc} 
\newcommand{\ran}{\texttt{ran}}
\newcommand{\ranb}[2][]{\ran_{#1} \! \glb #2 \grb}  

\newcommand{\loga}[2][]{\log^{#1}\! \gla #2 \gra}  
\newcommand{\logb}[2][]{\log^{#1} \glb #2 \grb}  

\newcommand{\prob}{\mathbb{P}}

\newcommand{\nran}[2]{\FUNCTION[#1,#2]{nran}}

\newcommand{\gla}{\,}  
\newcommand{\gra}{}  
\newcommand{\glb}{\left(}  
\newcommand{\grb}{\right)}  
\newcommand{\glc}{\left[}  
\newcommand{\grc}{\right]}  
\newcommand{\gld}{\left\{}  
\newcommand{\grd}{\right\}}  

\newcommand{\TO}{,\ldots,}
\newcommand{\VEC}[1]{\mathbf{#1}}
\newcommand{\xvec}{\VEC{x}}
\newcommand{\Xvec}{\VEC{X}}

\newcommand{\sigmavec}{\boldsymbol{\sigma}}
\newcommand{\tauvec}{\boldsymbol{\tau}}
\newcommand{\Upsilonvec}{\boldsymbol{\Upsilon}}

\newcommand{\mean}[1]{\left\langle #1 \right\rangle}
\newcommand{\half}{\frac{1}{2}}

\newcommand\bigOb[1]{\ensuremath{\OCAL\glb #1 \grb}}

\def\keyFont{\fontsize{8}{11}\helveticabold }
\def\firstAuthorLast{Hukushima {et~al.}} 
\def\Authors{Koji Hukushima\,$^{1,*}$, Werner Krauth\,$^{2,3,4}$
}
\def\Address{$^{1}$ Graduate School of Arts and Sciences,
    The University of Tokyo,
    Komaba, Meguro-ku, Tokyo 153-8902, Japan\\
$^{2}$ Laboratoire de Physique de l’Ecole normale
supérieure,
ENS, Université PSL, CNRS, Sorbonne Université, Université Paris Cité,
 Paris, France \\
$^3$ Rudolf Peierls Centre for Theoretical Physics, Clarendon
 Laboratory,
 University of Oxford, Oxford OX1 3PU, UK \\
$^4$ Simons Center for Computational Physical Chemistry,
New York University, New York (NY), USA
}

\def\corrAuthor{Koji Hukushima}

\def\corrEmail{k-hukushima@g.ecc.u-tokyo.ac.jp}

\begin{document}
\onecolumn
\firstpage{1}

\title[Damage spreading and coupling]{Damage spreading and coupling in spin
glasses and hard spheres}

\author[\firstAuthorsLast]{\Authors} 
\address{} 
\correspondance{} 

\extraAuth{}%

\maketitle


\begin{abstract}
We study the connection between damage spreading, a phenomenon long discussed in
the physics literature, and the coupling of Markov chains, a technique used
to bound the
mixing time. We discuss in parallel the Edwards-Anderson spin glass model
and the hard-disk system, focusing
on how coupling provides insights into the region of fast coupling within the
paramagnetic and liquid phases. We also work out the connection between path
coupling and damage spreading. Numerically, the scaling analysis of the mean
coupling time determines a critical point between fast and slow couplings. The
exact relationship between fast coupling and disordered phases has not been
established rigorously, but  we suggest that it will ultimately enhance our
understanding of phase behavior in disordered systems.

\tiny
  \keyFont{ \section{Keywords:} Spin glasses, hard-sphere model,
Markov chains, coupling times, damage spreading, thermodynamic phase
transitions, dynamic phase transitions}
\end{abstract}

\section{Introduction}

Monte Carlo simulations based on Markov chains~\cite{LandauBinderBook2013,SMAC}
play an
important role in the study of complex systems in physics and other sciences. In
a given
sample space, Markov chains perform random walks that, in their large-time
steady state, visit configurations according to a prescribed stationary
distribution (often the Boltzmann distribution). At early times, in contrast,
after its start from a given initial configuration, each Markov chain samples
different time-dependent distributions. The characterization of convergence
(that is, of the mixing timescale~\cite{Levin2008} for approaching the
stationary distribution)
is of greatest importance as, by definition, convergence is required for
sampling from the prescribed distribution and for estimating mean values of
observables (pressure, specific heat, internal energy) as running averages.
Moreover, the mixing timescale by itself carries important information on the
sampling problem. In a physics context, the sudden slowdown of mixing and
relaxation times (without any reference to observable) often
indicates a phase
transition. Well-known examples are the slowdown of the Glauber dynamics at the
paramagnetic--ferromagnetic transition in the Ising
model~\cite{Martinelli1999,Dyer2004}, as well as
the glass
transition, which is defined through the slowdown of relaxation processes
(although it is not of thermodynamic origin). Furthermore, the spin-glass
transition is believed to be signaled by a stark increase
of the relaxation times at low temperatures~\cite{EdwardsAnderson1975}.
Also, in certain local
Monte Carlo
algorithms for particle systems, fast mixing (in a way that we will discuss
later) is only possible in the liquid phase~\cite{Helmuth2022}, so a statement
about thermodynamic phases  is obtained from an analysis of mixing times,
without invoking observables. However, establishing mixing and
relaxation times can be an arduous task, both in practice and in
theory.

\begin{figure}[]
 \centering
\includegraphics[align=t,
width=0.31\textwidth]{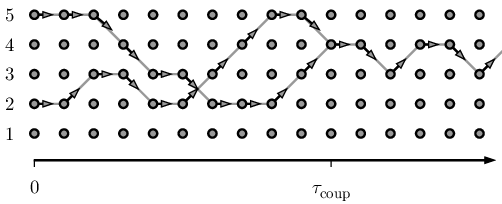}
\includegraphics[align=t,
width=0.31\textwidth]{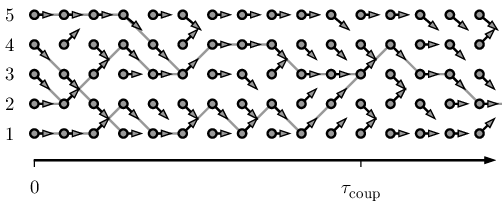}
\includegraphics[align=t,
width=0.31\textwidth]{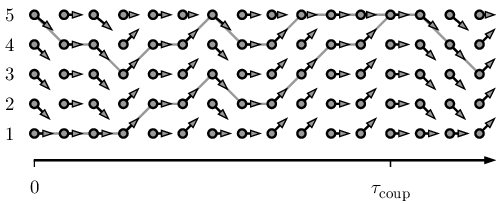}
 \caption{
Coupling for the random walk on a path graph (arrows point into the
three directions with equal probabilities, those leaving the graph are replaced
by straight arrows).
\subcap{Left} Classic coupling: the two random walks advance independently until
they merge at $\taucoup$.
\subcap{Middle} A random-map implementation of the classic coupling (independent
arrows).
\subcap{Right} A \standard, monotone coupling. Trajectories cannot cross.
}
\label{fig:Coupling}
\end{figure}

As convergence sets in, samples and empirical mean values (running averages)
become independent of initial configurations. Much stronger than mere
independence, samples can actually become identical for two (or more) different
initial configurations. This phenomenon, called coupling, is a focus of the
present paper. A coupling is a bivariate stochastic process that starts from two
far-away initial configurations at time $t=0$, say, $x_0$ and $y_0$, under the
condition that the projected evolution of $x_t$ and of $y_t$, taken separately,
realize a Markov chain with its transition matrix $P$. When the evolutions
of the two trajectories meet at the coupling time $\taucoup$, with
$x_{\taucoup} = y_{\taucoup}$, they are glued together for all later times (see
lhs of \fig{fig:Coupling}). Couplings of a given Markov chain can take many
different forms, but for all of them, the coupling time provides an upper bound
for the mixing time. This property has been used for almost a century to
 prove theorems on Markov chains \cite{Doeblin1938}, as cited
in \REF{Griffeath_MaxCoup1975}. Among many other developments, a more recent
version of coupling, known as \quot{coupling from the past}~\cite{ProppWilson1996},
has allowed for the perfect sampling of the stationary distribution without any
error, completely sidestepping the estimation of mixing time scales.

The path-coupling
approach~\cite{BubleyDyer1997}
attempts to  bound the \emph{global} coupling time through an
analysis which is \emph{local}
in both time and space.
The two far-away initial
configurations are imagined as end points of a \quot{path}
of many configurations. Configurations that are connected on the path are
neighbors
in the sample space with respect to a given metric. For the
one-dimensional random walk, the metric may correspond
to the Euclidean distance (see the lhs of \fig{fig:Coupling}).
For Ising systems,
the metric could be the Hamming distance: neighboring configurations differ
by only one spin. Similarly, for low-density systems of $N$
hard spheres, neighboring
configurations differ in only one sphere, which can be arbitrarily far away in
the two configurations, while the other
$N-1$ spheres coincide. It is often possible to deduce upper limits
for the coupling time
from the  contraction rates for the individual
path links. Path coupling was foreshadowed in the physics literature in a
phenomenon termed \quot{damage spreading}~\cite{Stanley1987}, which also
studied such neighboring
configurations under coupled-Markov-chain dynamics, a special type of coupling
for Glauber dynamics.
In the Ising model, for the
same dynamics, the damage was found to disappear rapidly
throughout the paramagnetic phase, a phenomenon later understood through the
concept of \quot{monotone coupling}.
In the Ising spin glass model, the damage was
found to disappear above a finite temperature in the paramagnetic phase, even
in two spatial dimensions, where the spin-glass transition temperature
is believed to vanish. Attempts to directly connect the damage spreading with a
thermodynamics process, for example a percolation transition, were finally
unsuccessful.
The connection between damage spreading, in other words path coupling, and the
thermodynamics is that \quot{fast} path coupling implies fast coupling, which
implies fast mixing. Fast mixing, in turn,  very often implies, in a physics
context, that the thermodynamic phase is trivial. This can lead to non-trivial
rigorous
bounds on the extension of the paramagnetic phase for spin
models~\cite{Dyer2004} or the
liquid phase for particle systems~\cite{Helmuth2022}.

This paper presents a unified description of coupling and of damage
spreading, using spin-glass and hard-sphere models as examples. In
\sect{sec:Foundations}, we provide common definitions, discuss theoretical
foundations, and explore the connection between coupling and mixing,
as well as the relationship between
the aforementioned path coupling and damage spreading. We also introduce the
scaling approach to phase transitions that we later apply to the coupling
phenomenon. \SECT{sec:SpinGlasses} is dedicated to spin glasses. We discuss
rigorous results and the generally accepted theoretical framework for the
spin-glass model introduced by Edwards and Anderson.
Additionally, we explore path coupling
and damage spreading for this model.
We further apply the scaling analysis to its mean coupling time,
which suggests a phase transition between fast and slow couplings.
\SECT{sec:HardSpheres} addresses the hard-sphere
model, for which we can generally transpose all the theoretical approaches of
\sect{sec:SpinGlasses}. The conclusions of our work are presented in
\sect{sec:Conclusions}.

\section{Theoretical foundations}
\label{sec:Foundations}

In this section, we discuss some fundamentals of Markov chains, and first
concentrate on the connection between the convergence of a Markov chain
expressed through its mixing time and any of its  couplings
(\subsect{subsec:FoundationsCoupling}). The special case of \quot{monotone}
coupling, which we also address, has important consequences for the
ferromagnetic Ising model, although it does not apply to the spin-glass models
or to hard spheres in more than one
dimensions~\cite{RandallWinklerInterval2005}.
We then discuss damage spreading in terms of path
coupling (\subsect{subsec:FoundationsDamagePathCoupling}). We will discuss the
intimate relation between a global view on coupling and a purely local view,
which only surveys configurations that differ minimally. We finally discuss in
\subsect{subsec:FoundationsScaling} the scaling approach to coupling that later
will be shown to apply both to spin glasses and to hard spheres.

\subsection{Mixing, coupling, and monotone coupling}
\label{subsec:FoundationsCoupling}

We consider a Markov chain with samples $x_t$, at time $t=0,1, \dots$ in a
sample space $\Omega$. In our case, its transition matrix $P$ implements
the heat-bath algorithm~\cite{Glauber1963,Creutz1980HeatBath,Geman1984} (in
other words, Glauber dynamics) for the
\EA or a version of the Metropolis
algorithm~\cite{Metropolis1953JCP} for hard spheres. We define the element
$P(x,x')$ as the
conditional probability to move from configuration $x$ at time $t$ to
configuration $x'$ at time $t+1$.
With an initial configuration $x_0$, the
distribution $\pit{t=0}$ is a Delta function centered at $x_0$.
The distribution evolves over time as
$\pit{t+1}(x') = \sum_x \pit{t}(x) P(x,x')$ for each time step $t$.
The approach to equilibrium is quantified by
the mixing time, which is the time it takes for $\pit{t}$ (which depends on the
choice of $x_0$) to approach the stationary distribution $\pit{t \to \infty} =
\pi$:
\begin{equation}
\taumix(\epsilon) = \min_t \Big \{
\underbrace{\max_{x_0\in\Omega}\left\|\pit{t}
 -\pi\right\|_\text{TV}}_{d(t)}<\epsilon \Big \}.
\label{equ:TVD_Definition}
\end{equation}
Here,
$\|\cdots\|_\text{TV}$ denotes the total variation distance~\cite{Levin2008},
that is, one
half of the absolute difference between $\pit{t}$ and $\pi$ over all the sample
space, and $\epsilon $ is an arbitrary positive parameter that has to be taken
smaller than $\half$. In \eq{equ:TVD_Definition}, the
\quot{$\max$} refers to
the worst initial choice for the approach of $\pit{t}$ (which depends on
$x_0$) to $\pi$, and this allows one to define the distance $d(t)$ between
$\pit{t}$ and $\pi$, without explicit reference to the starting distribution
$\pit{t=0}$. The mixing time is a non-asymptotic time
scale~\cite{AldousDiaconis1986} that describes the
initial approach of $\pit{t}$ towards the equilibrium distribution $\pi$, on a
finite distance scale $\epsilon$. It comes with an exponential bound, valid
from $\taumix$ up to $t \to \infty$, while the asymptotic approach towards
equilibrium, described by the (absolute) inverse gap of the transition matrix,
can be much faster~\cite{Levin2008}.

For a given transition matrix $P$ of a Markov chain on a sample space
$\Omega$, a coupling is defined
as a bivariate stochastic process with a configuration $(x_t , y_t)$ at time $t$
on the sample space $\Omega \times \Omega$, such that:
\begin{align}
\prob\glc x_{t+1} = x'\ | \ (x_t, y_t) = (x, y) \grc  &=  P(x, x'),
\\
\prob\glc y_{t+1} = y'\ | \ (x_t, y_t) = (x, y) \grc  &=  P(y, y').
\end{align}
The bivariate process that updates the two copies $x$ and $y$ need not be
Markovian~\cite{Griffeath_MaxCoup1975}, at a difference of its two projections.
Non-Markovian couplings are
of theoretical importance, but have not been used yet in applications.
Markovian couplings are described by a
transition matrix $\Pcoup{\cdot}{\cdot}$ on the sample space $\Omega \times
\Omega$, that satisfies:
\begin{align}
\sum_{y'}\Pcoup{x,y}{x', y'}  = P(x, x'),
\\
\sum_{x'}\Pcoup{x,y}{x', y'}  = P(y, y'),
\end{align}
so that the transition matrix of the coupled Markov chain, which acts on two
copies of the sample space $\Omega$, when projected on either copy,
gives back the original transition matrix.

Couplings can take a variety of forms. The \quot{classic} coupling performs two
statistically independent Markov chains until, by
accident, they couple, from when on they are glued together:
\begin{equation}
\Pcoup{x, y}{x',y'} =
\begin{cases}
P(x, x') P(y, y')
& \text{if $ x \neq y$},  \\
P(x, x')
& \text{if $ x = y, x' = y'$},  \\
0                 & \text{if $x = y, x'\neq y'$},
\end{cases}
\label{equ:ClassicCoupling}
\end{equation}
(see lhs of \fig{fig:Coupling}). At the coupling time $\taucoup$, the
trajectories first meet:
\begin{equation}
 \taucoup = \min_t\gld   x_t = y_t \grd.
\end{equation}

Transition matrices, as the ones in \eq{equ:ClassicCoupling}, are implemented
in Monte Carlo algorithms with the use of random elements, that is, one or
several random numbers for selecting a particle or a spin, for choosing a move,
and for accepting or rejecting it, etc.
For example, the move
from $x$ at time $t$ may produce an outcome $x'$ that depends on the
realization of the random element, but when this element is specified, as
$\Upsilon_t(x)$, it becomes a function, called random
map $ \SET{t} \times \Omega \to \Omega:
x \to x' = \phi\glc x, \Upsilon_t(x)\grc$.
The random
map $\phi\glc x, \Upsilon_t(x)\grc$ implementing this move must satisfy
\begin{equation}
\prob\gld \phi[x, \Upsilon_t(x)]  = x' \grd = P(x, x'),
\end{equation}
as it must
reproduce the
transition matrix $P$. A random map $\phi$ also specifies a coupling, and it
automatically implements
\quot{gluing}
operation, as two Markov chains that meet at a position $x$ at
time $t$ encounter the same random element.
For the classic coupling of \eq{equ:ClassicCoupling}, the randomness at time
$t$ is a vector $\Upsilonvec_t = \SET{\Upsilon_t(x): \ x \in \Omega}$ of
i.i.d random variables, that is of random numbers drawn from the same
distribution (see center of \fig{fig:Coupling}). For the \quot{\standard}
coupling, one uses, at time $t$,  the same random element
for all configurations $x \in \Omega$:
$\Upsilonvec_t = \SET{\Upsilon_t \TO \Upsilon_t} $.
Many other couplings exist, and
it is only of importance that the projection onto a single copy produces a
valid Markov chain.
While every random map corresponds to a coupling, it appears that not all
couplings (as for example the path couplings in \REF{Helmuth2022}) can be
expressed as random maps.

The connection between mixing times and coupling times is as
follows~\cite[corollary 5.3]{Levin2008}:
\begin{equation}
 d(t) \le \max_{x_0, y_0 \in \Omega} \prob_{x_0,y_0} \gld \taucoup > t \grd,
 \label{equ:MixingCoupling}
\end{equation}
where $d(t) $ is the distance entering the definition of the mixing time
in \eq{equ:TVD_Definition}. From our previous discussion, it is evident that
for random walks on large graphs, the classic coupling time can be much larger
than the mixing time, simply because the two Markov chains must hit the same
configuration at the same time. In contrast, the \standard coupling time is of
the same order as the mixing time for many random walks. In the problems at the
focus of this paper, we will witness different regimes, as a function of
external parameters, that are separated by a phase transition. In this context,
it is of great interest that an optimal coupling~\cite{Griffeath_MaxCoup1975}
realizes the coupling at time $t$ and at position $x_t$ of two Markov chains
that have started at time $t = 0$ at conﬁgurations $x_0$ and $y_0$ with the
minimum of the probabilities to go from $x_0$ or from $y_0$ to $x_t$ . The
optimal coupling is non-Markovian and virtually impossible to construct in
practice, but it demonstrates that  the bound of \eq{equ:MixingCoupling} can be
saturated.

A special class of couplings for which the inequality of
\eq{equ:MixingCoupling} can be tight (up to logarithms), requires the concept
of monotonicity.
In monotone couplings, there exists a partial ordering \quot{$\preceq$}
of
configurations so that $x_t \preceq y_t$, implies $x_{t+1} \preceq y_{t+1}$. In
terms of the random map, $x \preceq y$ implies
$\phi(x, \Upsilonvec) \preceq \phi(y, \Upsilonvec)$. For the
random walk on the path graph with a classic coupling, no partial ordering
exists, and trajectories of Markov chains may cross (see
rhs of
\fig{fig:Coupling}).
In
contrast, for the \standard coupling of the one-dimensional random walk, the
ordering is complete.
For a monotone coupling, with $l$  the length of the longest \quot{chain} in the
partially ordered subset the mean coupling time $\taucoup$  satisfies
\begin{equation}
\mean{ \taucoup}  < 2 \taumix(1/e) (1 + \loga{l}).
\label{equ:CouplingUpperBound}
\end{equation}
With \eq{equ:MixingCoupling}, there are thus upper and lower bounds for the
monotone coupling time in terms of the mixing time, and the two agree up to a
logarithm. For a monotone coupling with extremal elements, one must only survey
their evolution, that will bracket all other configurations  (see rhs of
\fig{fig:Coupling}). Full surveys are possible in other
cases~\cite{ChanalKrauth2008}, but the upper bound in
\eq{equ:CouplingUpperBound} is then often lost.

\subsection{Path coupling and damage spreading}
\label{subsec:FoundationsDamagePathCoupling}

We can consider families of Markov chains, that correspond to physical systems
with size $N$, which may represent the number of sites, of spins, or
of particles. As $N$ increases and approaches infinity, under suitable
conditions, such as constant temperature for spin systems or constant density
for particle systems, the behavior of these systems can be studied.
We may refer to \quot{fast} coupling if
the mean coupling time  $\mean{\taucoup}$
scales not
slower than a power of the system size
$N$ (in later sections, we will use an $N \loga{N}$ scaling).

As mentioned in the introduction, we may imagine the worst-case initial
configurations $x_0 $ and $y_0$ as the end points of a path
of configurations, with adjacent elements on the path being neighbors, with
respect to some metric. Under some conditions, it is often possible
to show that any pair of neighboring configurations come in expectation
even closer after one step of the Markov chain, and this establishes
that the distance between $x_1$ and $y_1$ contracts, and similarly for
later times, leading to a proof of fast coupling~\cite{BubleyDyer1997}.

The path-coupling analysis, that is local in sample space and in
time, yet valid uniformly for any
pair of neighboring configurations, yields a rigorous global fast-coupling
bound. We will discuss the limiting
temperature $\Tpath$ for spin glasses and limiting density $\etapath$ for
hard sphere systems, for
which the uniform contraction allows one to prove fast coupling.
However, the path-coupling approach is quite conservative. Numerical
evidence~\cite{BernardChanalKrauth2010} indicates fast coupling
down to a temperature $\Tcoup$ which is lower than $\Tpath$, and up to a density
$\etacoup$ which is higher than $\etapath$. However, only $\Tpath$ and
$\etapath$ are known analytically. In the models that we study, the
coupling is either exponential (and thus \quot{slow}) or \quot{fast}.

The path-coupling analysis provides a justification for \quot{damage spreading},
which was much studied for spin systems in the physics literature, with the
\standard coupling. As in path coupling, two neighboring initial configurations
$x_0$ and $y_0$ were chosen, but they were followed up to very large times. The
explicit relation between the time to couple and the time to mix is lost, but
the mean coupling time starting from neighboring initial configurations is again
exponential below $\Tcoup$ and $\sim N \loga N$ or faster above.
The connection between coupling and damage spreading was made in
\REF{BernardChanalKrauth2010}.

\subsection{From rigorous to non-rigorous approaches to coupling, scaling
approach results}
\label{subsec:FoundationsScaling}

The coupling time in \eq{equ:MixingCoupling} that allows to bound the mixing
time follows the
worst-case pair of starting configurations $x_0$ and $y_0$. For monotone
coupling, these configurations are given by the two extremal elements,
but in general, this
requires a survey of the entire sample space. For the Glauber dynamics of spin
glasses with the \standard coupling, the patch
algorithm~\cite{ChanalKrauth2008} rigorously surveys the
$|\Omega| \sim 10^{600}$ configurations on a $64\times 64$ lattice, and the same
algorithm also applies to hard-sphere
models~\cite{ChanalKrauth2010,BernardChanalKrauth2010}. It was found, however,
that a few hundred random initial configurations contained worst-case pairs
with high probability. Such a partial-survey approximation is easy to
set up in practice.

We use the partial-survey approximation to evaluate the mean coupling time $\mean{\taucoup}$
for spin-glass and hard-sphere systems.
Here, a systematic numerical approach, inspired by finite-size scaling
analyses of second-order phase transitions, is discussed for distinguishing
between fast and slow couplings.
In this context, fixing the system size $N$ corresponding to limiting
the worst-case pair distance between initial configurations, and by varying
$N$, the scaling behavior is analyzed as $N$ grows.
Suppose we obtain $\mean{\taucoup}(N,\beta)$ numerically
as a function of the system size $N$ and the model parameter $\beta$, which
represents the inverse temperature
in the case of spin-glass systems. For hard-sphere systems, this parameter may
also be the density $\eta$. In the fast-coupling regime,
the size dependence of $\mean{\taucoup}$ exhibits $N\log N$ behavior at high temperature,
while in the slow-coupling regime, it increases exponentially at low temperatures.
This phenomenon can be viewed as a dynamical phase transition,
with the two behaviors changing at a certain critical temperature
$\beta_\mathrm{coup}$.

Assuming that, as $\beta$ approaches $\betacoup$, $N^*(\beta)$ provides
a diverging scale that controls the coupling behavior,
the scaling form is postulated to hold in the
vicinity of $\betacoup$, expressed as
\begin{equation}
  \mean{\taucoup}(N,\beta) = N^\phi f(N/N^*(\beta)) \ \   \mathrm{with}
\ \ N^*(\beta) =|\beta_\mathrm{coup}-\beta|^{-\omega},
\label{equ:fss}
\end{equation}
where $\phi$ and $\omega$ are positive parameters associated with the dynamical
transition, and $f$ is a universal scaling function.
The two behaviors of fast and slow couplings are represented
in the asymptotic form of this
scaling function $f(x)$, with $x=N|\beta_\mathrm{coup}-\beta|^\omega$:
\begin{equation}
 f(x) =
 \begin{cases}
   x^{1-\phi}\loga{x} &  \text{as\ } x\to\infty \text{\ \ \ for \ \ \ }
\beta_\mathrm{coup}>\beta,\\
  \expb{a x} & \text{as\ } x\to\infty  \text{\ \ \ for \ \ \ }
\beta_\mathrm{coup}<\beta,
 \end{cases}
\end{equation}
with a positive constant $a$.
The value of the scaling function $f(0)$ at $\beta=\betacoup$ is constant,
and the parameter $\phi$ can be identified as the exponent of the power-law
divergence of $\taucoup$ at $\betacoup$.
In the case of a ferromagnetic Ising model with monotone coupling,
where the coupling time and the mixing time coincide,
these parameters characterize the universality class of the corresponding
ferromagnetic phase transition and are related to the dynamical exponent $z$
and the correlation length exponent $\nu$ through the dimensionality $d$.
For example, in the case of the mean-field ferromagnetic Ising model,
it has been rigorously shown that $\phi=3/2$\cite{Ding2009},
which is consistent with $z=2$.
However, in general, the singularity at $\betacoup$ in this coupling time
is not directly associated with an order parameter of the physical system.

\section{Coupling in spin glasses}
\label{sec:SpinGlasses}

This section examines the coupling in the
\EA~\cite{EdwardsAnderson1975} of spin glasses, focusing on the dynamical
properties of its Glauber dynamics. We first review known exact results on the
thermodynamics of the model  in finite dimensions
(\subsect{subsec:SpinGlassesRigorous}), followed by an analysis of path coupling
and numerical calculations (\subsect{subsec:SpinGlassesPathScaling}). Finally,
we discuss the physical significance of these findings
(\subsect{subsec:SpinGlassesDendrograms}).

The \EA describes
$N$ Ising spins $\sigmavec = \SET{\sigma_0 \TO \sigma_{N-1}}$
with $\sigma_k = \pm 1$ on a $d$-dimensional hypercubic lattice
with periodic boundary conditions  and even side length $L$.
The stationary weight $\pi(\sigmavec)$ of each configuration is given through
its energy $E(\sigmavec)$ as follows:
\begin{equation}
 \pi(\sigmavec) = \expc{ - \beta E(\sigmavec)}\quad
 E(\sigmavec) = -\sum_{\langle ij\rangle}  J_{ij} \sigma_i \sigma_j,
 \label{equ:BoltzmannSpin}
\end{equation}
where $\mean{ij}$ denotes the sum over nearest-neighbor pairs of spins. For each
spin-glass sample, the interactions $J_{ij} = J_{ji} \in \SET{-1,+1}$ are
quenched (that is, fixed). The ensemble average is obtained by taking
the $J_{ij}$ as i.i.d., with $J_{ij} = +1$ or $J_{ij} = -1$
with equal probability. In our
statements about mixing and coupling, this ensemble average is understood.

We consider two versions of the heat-bath algorithm, namely random updates
and parallel updates. For the random updates, at each time step,
starting from a configuration $\sigmavec(t)= \SET{\sigma_0 \TO \sigma_{N-1}}$,
one random spin $\sigma_k$ among the $N = L^d$ spins is sampled. At
time $t+1$,
the configurations
$\sigmavec^+ = \SET{\sigma_0 \TO \sigma_{k-1}, +1, \sigma_{k+1} \TO \sigma_{N-1}}$
and
$\sigmavec^- = \SET{\sigma_0 \TO \sigma_{k-1}, -1, \sigma_{k+1} \TO \sigma_{N-1}}$
are chosen with probability $\pi(\sigmavec^+)
/\glc \pi(\sigmavec^+) + \pi(\sigmavec^-) \grc$
and $\pi(\sigmavec^-) /\glc \pi(\sigmavec^+) + \pi(\sigmavec^-) \grc$,
respectively. These probabilities can be written as $\pi^+(h_k)$
and $1-\pi^+(h_k)$, through the
local field $h_k = \sum_{j \in \text{nbr}(k)} J_{kj} \sigma_j$,
with the sum over the neighboring sites $j$ of site $k$.
For parallel updates, on a bipartite lattice, as the hypercubic lattice with
even $L$,
the energy couples term on different sub-lattices. In one
Monte Carlo cycle, first all the spins are updated on one sublattice, followed
by those on the other sublattice.
For simplicity, we count time in terms of \quot{Monte Carlo cycles}, that is,
$N$ updates, also for the random update case.

The classic coupling of \eq{equ:ClassicCoupling}, applied to the
heat-bath algorithm with the random updates,
randomly chooses two spins $\sigma_k$ and $\tau_k$ in order to
independently update the configurations $\sigmavec_t$ and $\tauvec_t$, until
they meet. In terms of random maps, this requires $2 \times 2^N$ random numbers
at each time $t$, one to choose the spin, and one to update it, which is not
practical. It is evident that at all temperatures, including infinite
temperature, the coupling time is exponential in $N$, as the trajectories have
to accidentally meet.

For the \standard coupling, the heat-bath algorithm for the random update
uses a source of randomness $\Upsilonvec_t$ given by:
\begin{equation}
 \Upsilonvec_t =
\SET{k, \Upsilon} =
 \SET{\underbrace{\nran{1}{N}}_{\text{lattice site $k$}},
\underbrace{\ranb{0,1}}_{\text{heat-bath}}}.
\label{equ:RandomnessStandard}
\end{equation}
In short, the randomness $\Upsilonvec_t$ samples the lattice site $k$ to be
updated, as well as the random number used for the
heat-bath update. The random-maps function $\phi$ is then defined
for a given spin configuration $\sigmavec$ and the randomness $\Upsilonvec_t$
as follows:
\begin{equation}
 \phi(\sigmavec, \Upsilonvec_t): \sigma_k(t+1) =
 \begin{cases}
 1 & \text{if $\Upsilon < \pi^+\left(h_k(\sigmavec)\right) =
 \glc 1 + \expa{- 2 \beta h_k(\sigmavec)} \grc^{-1}$
},  \\
-1 & \text{else },
 \end{cases}
\label{equ:SpinHeatbathUpdate}
 \end{equation}
where the local field is $h_k = \sum_{j \in \text{nbr}(k)} J_{kj} \sigma_j$.
We note that
$\sigma_k(t+1) $ does not depend on $\sigma_k(t)$.

For the parallel update on a bipartite lattice, the randomness
$\Upsilonvec_t$ is given by:
\begin{equation}
 \Upsilonvec_t =
\SET{\Upsilon^0, \Upsilon^1 \TO \Upsilon^{N-1}} =
\SET{
\underbrace{\ranb{0,1}}_{\text{site $0$}},
\underbrace{\ranb{0,1}}_{\text{site $1$}} \TO
\underbrace{\ranb{0,1}}_{\text{site $N-1$}}
}.
\end{equation}
The update is performed in two half steps, on the two sub-lattices, as
described earlier. The coupling corresponding to
\eq{equ:SpinHeatbathUpdate} is monotone only for the ferromagnetic case
($J_{ij} = +1$), where larger local fields are produced by larger
neighboring spins  $\sigma_j$.

\subsection{Spin glasses: From rigorous results to numerical simulations}
\label{subsec:SpinGlassesRigorous}

From a mathematical perspective, the fact that the interactions $\{J_{ij}\}$ are
quenched random variables complicates the analysis with respect to uniform
interactions. The Sherrington--Kirkpatrick model~\cite{Sherrington_1975}, in
other words the \EA on a complete graph corresponding to its
infinite-dimensional limit, has been at the forefront of theoretical
developments in spin-glass research. This model undergoes a thermodynamic phase
transition separating a high-temperature paramagnetic phase from a
low-temperature spin-glass phase at an exactly known temperature. The existence
of this phase transition and the low-temperature properties were first
established using the replica method~\cite{SpinGlassBeyond} and later proven
rigorously~\cite{talagrand2003spin}.

Mathematically rigorous results for the \EA in finite dimensions are very few.
The difficulty in analyzing finite-dimensional systems arises, in part, from the
presence of Griffiths singularities. In systems with random interactions, local
regions may exhibit low probabilities but  strong correlations, leading to
anomalous singularities in the free energy and divergences in high-temperature
expansions. In a specific random system, the existence of this type of
singularity has been mathematically proven, and is known as the Griffiths
singularity~\cite{Griffiths1969}. This singularity emerges at the phase
transition temperature when the random interactions are assumed to be uniform.
In the \EA, the Curie temperature of the ferromagnetic Ising model (with all
$J_{ij}$ equal to $+J$) constitutes this Griffiths temperature. Despite these
difficulties, it has been proven that, at sufficiently high temperature, the
so-called Edwards--Anderson order parameter vanishes identically, and the
spin-glass susceptibility remains finite in short-range spin glass
models~\cite{Froehlich1984,Berretti1985}. This means that the high-temperature
phase is paramagnetic, although rigorous temperature bounds seem to be absent.
These temperature regions are far from the spin-glass transition temperature
$\TSpinGlass$ suggested by numerical simulations mentioned below. One expects
that a spin-glass phase cannot exist at temperatures higher than the Griffiths
temperature, so that the Griffiths temperature likely serves as an upper bound
for $\TSpinGlass$. However, this seems not to be a rigorous statement.

Local Markov-chain Monte Carlo simulations, mostly with the heat-bath algorithm,
were first to suggest a finite spin-glass transition temperature in three and
higher dimensions, and a zero transition temperature in two
dimensions~\cite{BhattYoung1985,Houdayer2001}.
While neither has been proven rigorously, the
fact that the ground state of the two-dimensional \EA can be computed in a time
polynomial in $N$\cite{Bieche1980,ThomasMiddleton2009} supports the conclusion
of a vanishing transition temperature in two dimensions. These conclusions, both
for three and higher dimensions as well as for two dimensions, were based on
estimates of spin-glass order parameters, which, in essence, test to which degree
the equilibrium running average of a given observable, such as the spin overlap
between replicated systems, becomes independent of two independent starting
configurations (see \REF[Eq. 4]{BhattYoung1985}). Another route to obtaining
information on spin glasses has consisted in studying the autocorrelation
functions of observables (e.g., the value of $\sigma_k(t)$). Early results
already pointed to a difference of the scaling behavior at late times~\cite[Fig.
7]{Ogielski1985}, \cite{OgielskiMorgenstern1985}, from which a
finite spin-glass transition temperature in the range $\TSpinGlass \simeq 1.10 -
1.14$ was inferred. Although no consensus has been reached on the nature of the
spin-glass phase, more recent studies have refined estimates of the
spin-glass transition temperature $\TSpinGlass$ in three dimensions, with
different estimates such as $\TSpinGlass =1.1019(29)$\cite{Janus2013} and
$\TSpinGlass = 1.109(10)$\cite{Hasenbusch2008},
which combine simulations for rather small
system sizes with empirical extrapolations to the thermodynamic limit.

Damage spreading in spin-glass systems was found as a dynamical anomaly in
early numerical simulations~\cite{Derrida1987DynamicalPT,Campbell1991}, which
showed that it occurs at temperatures higher than the spin-glass transition
temperature suggested by other studies. However, it remained unclear whether the
anomaly was related to the spin glass transition itself or to the Griffiths
singularity. The connection between damage spreading and coupling, which is the
focus of this paper, was recognized in \REF{BernardChanalKrauth2010}.

\subsection{From path coupling to scaling plots}
\label{subsec:SpinGlassesPathScaling}

In the finite-dimensional \EA, we now consider the \standard
coupling for the heat-bath algorithm of \eq{equ:SpinHeatbathUpdate}. To
establish coupling, we consider two arbitrary spin configurations as initial
states of the two Markov chains and apply the path-coupling argument of
\subsect{subsec:FoundationsDamagePathCoupling}.
The two configurations differ in at most $N$ sites, so that we can connect them
by a path of at most $N$ neighboring configurations that differ by one spin
only.

\begin{figure}[htb]
 \centering
\includegraphics[align=t, width=0.495\textwidth]{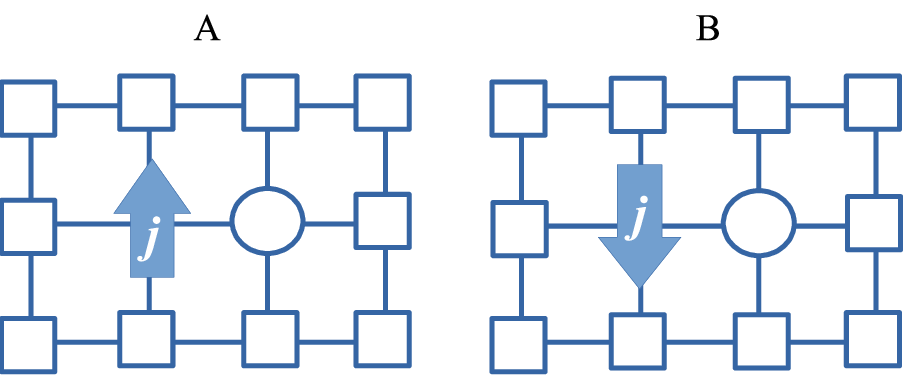}
\includegraphics[align=t, width=0.49\textwidth]{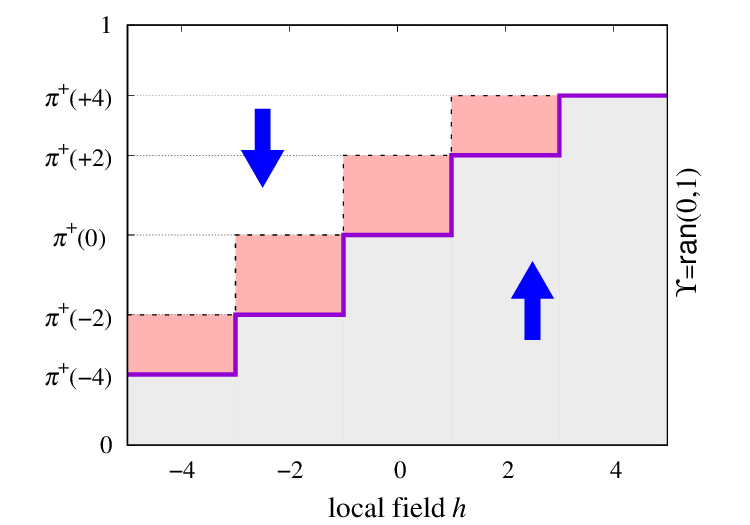}
 \caption{
\subcap{Left}
Two spin configurations, $\sigmaa$ and $\sigmab$, which differ at a single site
indicated by arrows. The sites connected to it are marked with squares and
circles, which represent arbitrary states, either up or down, that are common
to both configurations. \subcap{Right} The probability $\pi^+(h)$ of the next spin
state being ``up''($+$) in the heat-bath algorithm for a two-dimensional Ising
model, as a function of the local field $h$, following the form
$\pi^+(h)$ of \eq{equ:SpinHeatbathUpdate}. The next state becomes \quot{up} if a random
number $\Upsilon$ falls within the gray region. The red region represents
conditions where two spins, which differ by a local field of $2$, result
in different next states.
}
\label{fig:UpsilonHPlot}
\end{figure}

Let $\sigmaa$ and $\sigmab$ be two such neighboring configurations (see lhs of
\fig{fig:UpsilonHPlot}) that differ by the  spin $j$. The common random element
$\SET{k, \Upsilon}$ of \eq{equ:RandomnessStandard} contains the spin $k$ to be
updated and the random number $\Upsilon$ required for the heat-bath step of
\eq{equ:SpinHeatbathUpdate}. With probability $p_{1 \to 0} = 1/N$, the spin $j$
is updated. The field $h_j$ is the same for $\sigmaa$ and $\sigmab$, and so is
$\Upsilon$ in \eq{equ:SpinHeatbathUpdate}. It follows that the distance
decreases from $1 $ to $0$ with $p_{1 \to 0}$.

With probability $2d/N$, spin $l$, one of the $2d$ neighboring spins of $j$, is
updated. The local fields $h_l(\sigmaa)$ and $h_l(\sigmab)$ differ by exactly
$2$. The probability $p_{1 \to 2}$ of making different decisions, which
corresponds at most to the red region on the rhs of \fig{fig:UpsilonHPlot}, is
evaluated as:
\begin{equation}
 p_{1 \to 2} = \frac{2d}{N} \max_h \left| \pi^+(h)-\pi^+(h\pm 2)\right| =
 \frac{2d}{N} \glc
\pi^+(0) - \pi^+(-2) \grc =
 \frac{2d}{N}
 \glc
\half - \frac{1}{1 + \expb{4 \beta}} \grc.
\end{equation}
If $p_{1 \to 0 } > p_{1 \to 2}$, the expected
distance between $\sigmaa$ and $\sigmab$
decreases after one step, for any choice of spin configuration and any choice
of the couplings $\SET{J_{ij}}$, which is the case at high temperature. The
limiting temperature for the application of the path-coupling argument is when
$p_{1 \to 0 } = p_{1 \to 2}$, which translates into
\begin{equation}
 \betapath = \frac14 \logb{ \frac{2d}{d-1} -1} = \frac{1}{2d}
 + \frac{1}{6d^3 }
 + \frac{1}{10 d^5 }  + \cdots, 
\end{equation}
and equivalently, 
\begin{equation}
 \Tpath = \frac{1}{\betapath}=2 d - \frac{2}{3d }
 - \frac{8}{45 d^3 }  + \cdots.
\label{eqn:Tpath}
\end{equation}
For $T > \Tpath$, we are assured of fast coupling in the \EA. The argument also
holds for sub-lattice parallel updates. As discussed, $\Tpath$ is obtained for
any choice of interactions and any spin configuration. Consequently, $\Tpath$ is
also the path-coupling bound for the ferromagnetic Ising model, although we know
from monotonicity that fast coupling will take place down to the Curie
temperature.

\begin{figure}
\centering
 \includegraphics[width=0.49\textwidth]{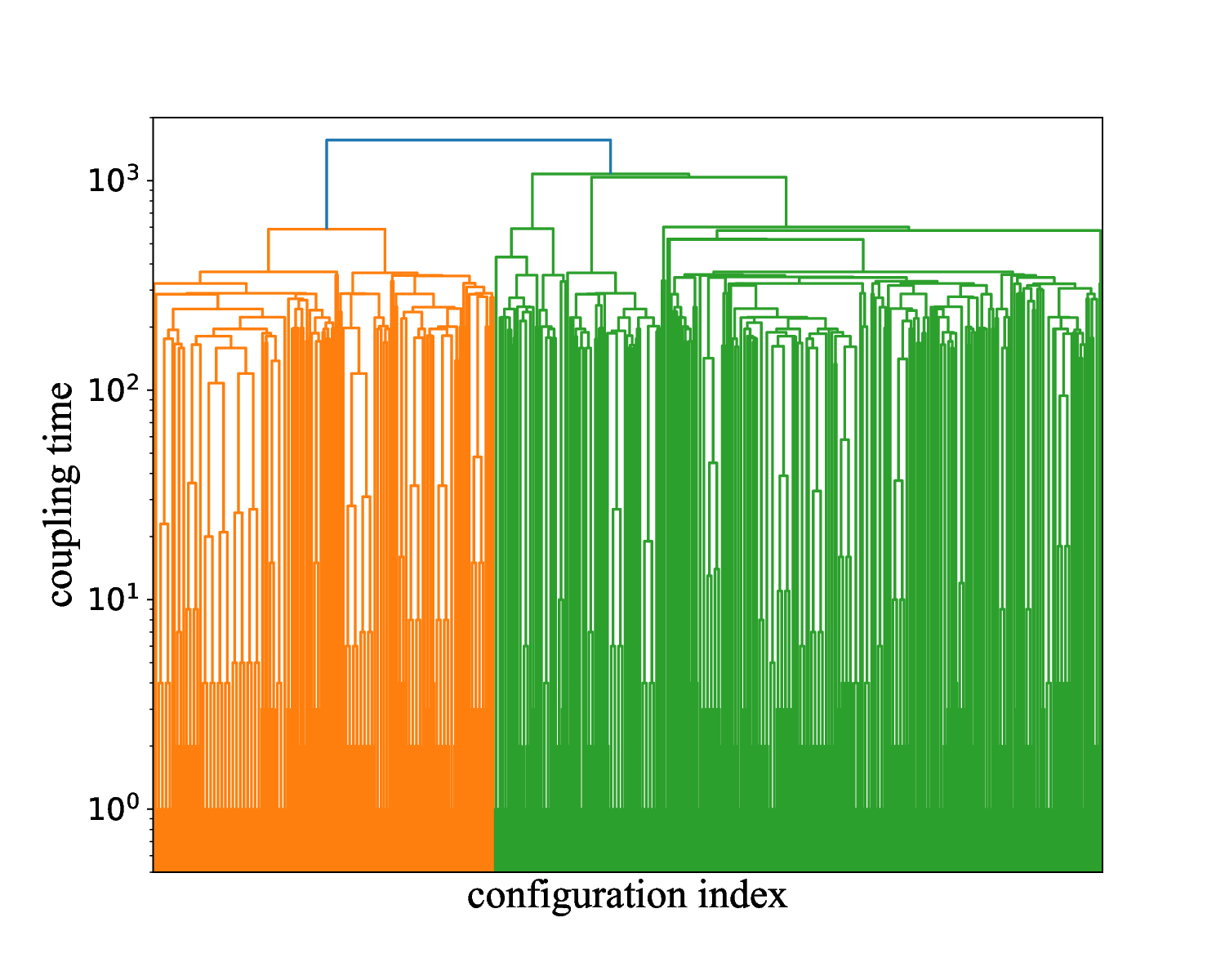}
 \caption{Dendrogram of configurations in the partial-survey approximation
for the  three-dimensional \EA with parallel updates
at $T =3.90 \gtrsim \Tcoup $ and with $\NCAL_0 = 512$.
Any starting set with representative
configurations in the two main branches (\emph{orange}, \emph{green})
gives the same coupling time, explaining the success of the partial survey.
}
\label{fig:Dendrogram}
\end{figure}

We now numerically evaluate the mean coupling time of the finite-dimensional
\EA,
in both two and three dimensions.
in view of the scaling analysis discussed in \sect{subsec:FoundationsScaling}.
The mean coupling time of the two-dimensional model was already evaluated under a
random update rule, and it has been demonstrated that a dynamical phase
transition occurs in which the size dependence of the coupling time
qualitatively changes~\cite{BernardChanalKrauth2010}, confirming earlier
results~\cite{Campbell1991}.
The mean coupling time results presented below are evaluated using the
partial-survey approximation with the number $\NCAL_0$ of randomly chosen
initial conditions.
The results obtained with different values of $\NCAL_0$ are plotted at
each data point, but they are completely contained within the size of the
markers, thereby confirming that they are independent of $\NCAL_0$.
A dendrogram representation explains the independence of the mean coupling
time of $\NCAL_0$ (see \fig{fig:Dendrogram}).

\begin{figure}
 \centering
\includegraphics[width=0.495\linewidth]{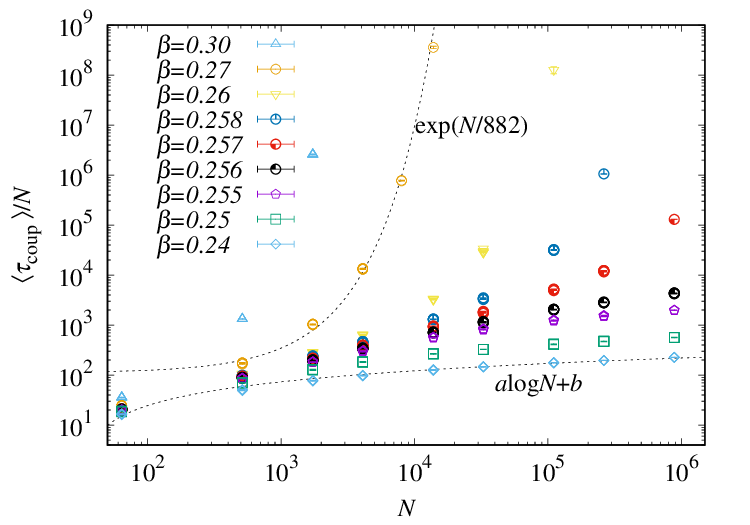}
\includegraphics[width=0.495\linewidth]{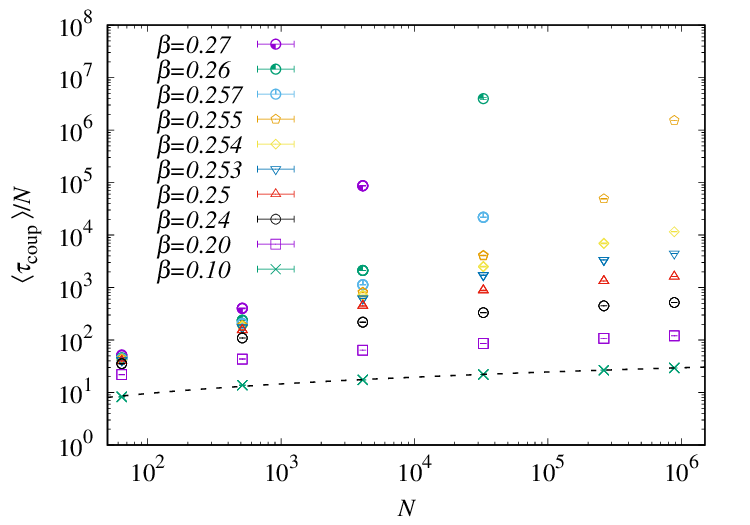}
 \caption{System-size $N$ dependence of the mean coupling time at various
inverse
temperatures in the three-dimensional \EA.
\subcap{Left}
Parallel
update. \subcap{Right} Random update.
}
\label{fig:Ctime_3dISG}
\end{figure}

The two panels in \fig{fig:Ctime_3dISG} show the estimated mean
coupling time for the partial-survey approximation under the parallel and random
updates in the three-dimensional \EA.
Although the two updates differ in the high-temperature limit, both exhibit a
$N\log N$ behavior for system size $N$ at sufficiently high but finite
temperatures. As the temperature decreases, the behavior of the $N$ dependence
of the mean coupling time changes from slow to fast increase at a certain
temperature. There is a slight, yet significant, difference in the transition
temperature between the two updates, with a lower transition temperature
observed for the parallel updates. This illustrates that coupling has no
direct
thermodynamic significance.

\begin{figure}
 \centering
\includegraphics[width=0.495\linewidth]{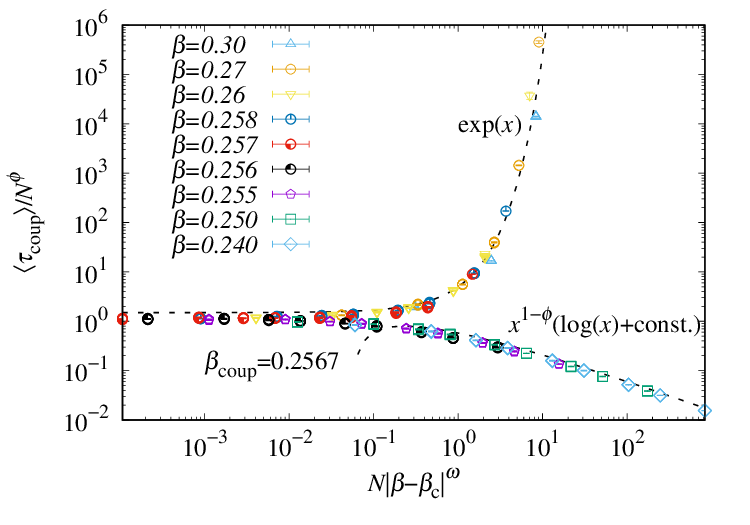}
\includegraphics[width=0.495\linewidth]{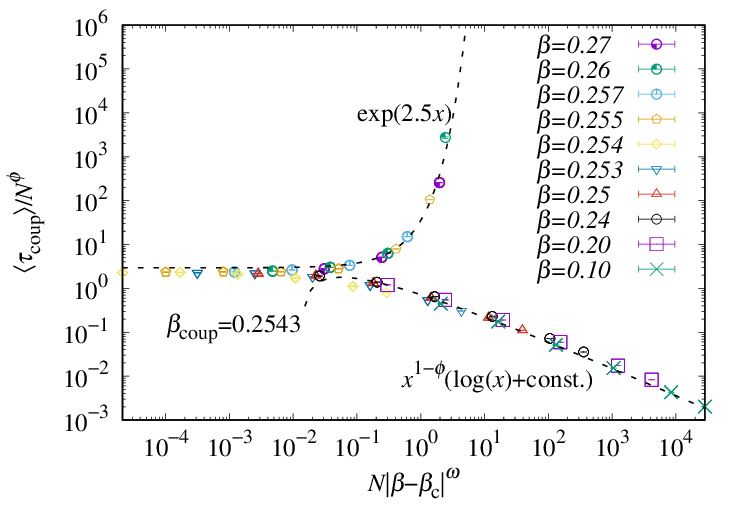}
\caption{
Finite-size scaling plot of the mean coupling time in the three-dimensional \EA.
\subcap{Left}
Parallel updates ($\omega \simeq 1.7$, $\phi \simeq 1.7$ and
$\beta_\mathrm{coup} \simeq 0.2567$).
\subcap{Right}
Random updates ($\omega \simeq 1.84$, $\phi \simeq 1.70$ and
 $\beta_\mathrm{coup} \simeq 0.2543$). Two dotted lines in each panel represent
the
expected high- and low-temperature asymptotic forms of the scaling function.
}

\label{fig:Ctime_3dISG-scaling}
\end{figure}

\figg{fig:Ctime_3dISG-scaling} presents finite-size scaling plots of the mean
coupling time for the three-dimensional \EA, comparing both the parallel and
random updates. The plot demonstrates that the scaling works well when the
appropriate scaling parameters are chosen. This is consistent with the above
argument that the transition temperatures, $\Tcoup$ or $\betacoup$,
are significantly different for the
two update rules. In contrast, the precision of the scaling exponents, $\phi$
and $\omega$, is not as precise as that of the transition temperature, and it
can be considered that these two rules yield almost the same values for these
exponents. It remains unclear whether these exponents have a meaning analogous
to the critical exponents of a second-order transition. Of particular interest
is the exponent $\omega$, which represents the divergence of the characteristic
scale as it approaches the transition temperature. Our results suggest that this
exponent has the same value on both the high and low temperature sides of the
transition temperature. This is comparable to the correlation length exponent.

\begin{figure}
 \centering
\includegraphics[width=0.495\linewidth]{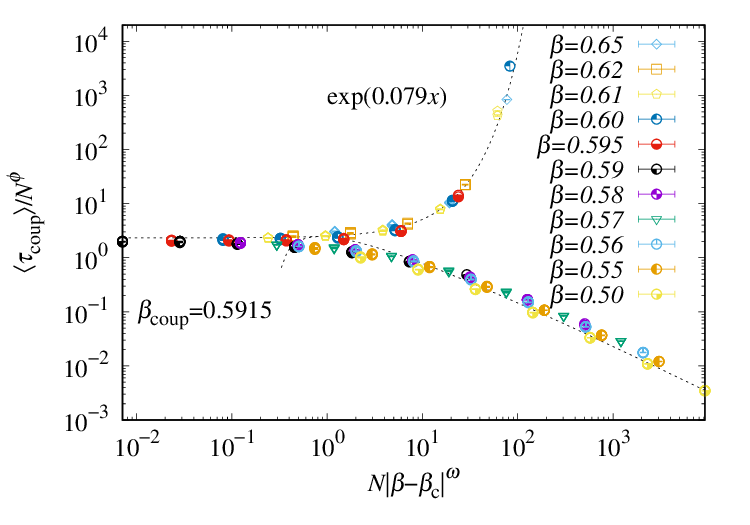}
\includegraphics[width=0.495\linewidth]{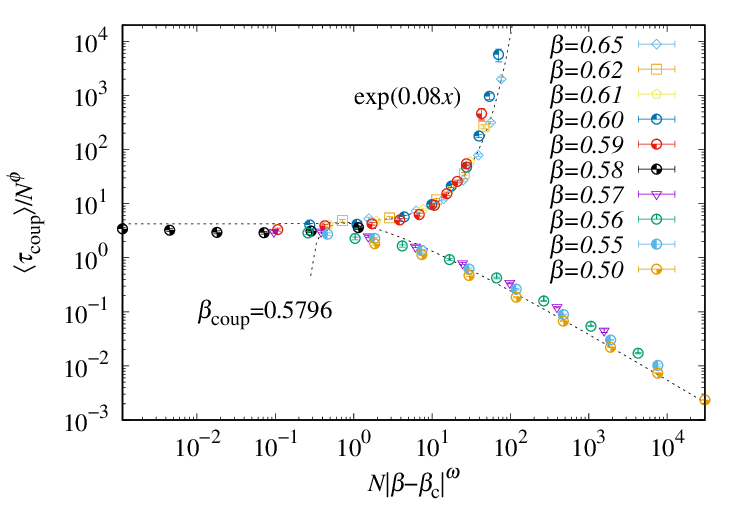}
\caption{
 Finite-size scaling plot of the mean coupling time in the two-dimensional \EA.
\subcap{Left} Parallel update (our simulations). The obtained
parameters are $\omega\simeq 1.4$, $\phi\simeq 1.95$ and $\beta_\mathrm{coup}\simeq
0.5915$.
\subcap{Right}
Random updates (original data of
\REF{BernardChanalKrauth2010}). The obtained parameters are $\omega\simeq 1.4$,
$\phi\simeq 1.95$ and $\beta_\mathrm{coup}\simeq 0.5796$.
Two dotted lines in each panel represent the expected high- and low-temperature
asymptotic forms of the scaling function.
}
\label{fig:Ctime_2dISG}
\end{figure}

An analogous scaling analysis for the two-dimensional \EA is shown in
\fig{fig:Ctime_2dISG}. The left panel is the analysis result of our own
numerical simulations using the sub-lattice parallel update, while the right
panel presents the scaling analysis based on numerical data using the random
update from \REF{BernardChanalKrauth2010}. In both cases, the scaling is
consistent with a phase transition in the mean coupling time. As observed in the
three-dimensional model, $\Tcoup$ depends on the underlying Markov chain, with a
lower transition temperature for the parallel update. The scaling exponents
depend on the dimensionality. However, the proper scaling variable may not be
the number of spins, $N$, used here, but rather the linear dimension $L$. This
suggests that the value of the exponents may depend on the dimensionality
through the relationship $N=L^d$.

\subsection{Path coupling and damage spreading for spin glasses}
\label{subsec:SpinGlassesDendrograms}

\begin{table}[]
    \centering
    \begin{tabular}{c|ccccc}
        Dimension $d$ & $\TSpinGlass$ & $\Tcoup$ (Parallel) & $\Tcoup$ (Random)
                 & $\TGriffiths$ & $\Tpath$  \\ \hline
        2 & $0$ & 1.69... & 1.72... & 2.269... & 3.640... \\
        3 & $1.1019 - 1.1090$ & 3.89... & 3.93...& 4.51...  & 5.770...
    \end{tabular}
     \caption{Spin-glass transition and coupling temperatures for the
     \EA in two and three dimensions.
$\TSpinGlass$ is the numerical estimate from \REF{Janus2013,Hasenbusch2008}
in three dimensions  and is expected~\cite{BhattYoung1988,Houdayer2001} to vanish in two dimensions.
$\Tcoup$ is from \figtwo{fig:Ctime_3dISG-scaling}{fig:Ctime_2dISG} and
$\TGriffiths$ is the Curie temperature of
the ferromagnetic Ising model. Finally, $\Tpath$ is from \eq{eqn:Tpath}.
}
\label{tab:spin-glass}
\end{table}

Table~\ref{tab:spin-glass} summarizes the key temperatures discussed
in previous sections, including $\Tpath$ and $\Tcoup$, as well as
previously estimated results for $\TSpinGlass$ and $\TGriffiths$.
This table demonstrates the differences in transition temperatures for both
two- and three dimensional \EA, providing a detailed overview of the coupling
and spin-glass transitions.

\begin{figure}[]
\centering
\includegraphics[align=t, width=0.49\textwidth]{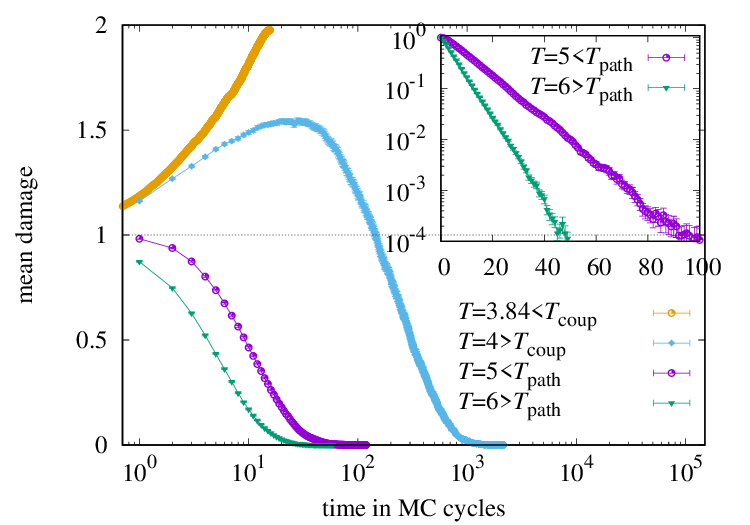}
\caption{
Damage evolution over time for two states differing by a Hamming distance of 1
as initial conditions in random updates of the three-dimensional
\EA.
The size is $N=64^3$, and the four temperatures shown are above
and
below both $\Tpath$ and $\Tcoup$. The inset shows the same plot
on a semi-log scale.
}
\label{fig:DamagePropagation}
\end{figure}

On the one hand, path coupling demonstrates that, above $\Tpath$,
the uniform contraction between neighboring configurations leads to fast
coupling. Below $\Tpath$, there are spins $k$
(for example, those with $h_k = 0$) for which at least initially, there is no
such contraction.
Nevertheless, as our numerical simulations show, fast $N \loga{N}$
coupling also takes place in the window $\Tcoup < T < \Tpath$. The absence of
a regime change at $\Tpath $ can be illustrated, in the language of damage
spreading, by following the mean damage as a function of time for two
configurations that initially, at time $t=0$, are neighboring. Above $\Tpath$,
the mean damage decreases exponentially for all times (see inset of
\fig{fig:DamagePropagation}), whereas for $T < \Tcoup$, it increases rapidly.
In
the window  $\Tcoup \lesssim T \lesssim \Tpath$, the mean damage initially
increases, as expected, but then turns around and again vanishes exponentially.
This turning point seems to occur when the damage reaches a certain size, which
grows as the temperature approaches $\Tcoup$. This behavior can be understood
in analogy with the characteristic diverging scale $N^*(\beta)$
in the finite-size scaling analysis of \eq{equ:fss}, which suggests
a picture similar to a critical phase transition, where the threshold damage
size corresponds to the diverging scale near $\Tcoup$.

\section{Coupling in hard spheres}
\label{sec:HardSpheres}

In this section, we examine coupling for the hard-sphere system of
statistical mechanics. For concreteness, we concentrate on the
two-dimensional hard-disk model, which was the object of the historically first
study using
Markov chains~\cite{Metropolis1953JCP}. The model has created an unabating
series of works in mathematics, physics, and chemistry~\cite{Li2022}. After an
introduction to the model and to the Metropolis
algorithm~\cite{Kannanrapidmixing2003} that we will mostly
consider, we review the very few known exact results on the model
(\subsect{subsec:HardSpheresRigorous}), and then move on to the analysis of path
coupling (\subsect{subsec:HardSpheresPathScaling}) and to numerical calculations
leading up to our scaling analysis. We finally discuss, following
\REF{Helmuth2022}, in which way the behavior of the algorithm teaches us about
the physics of the hard-disk model (\subsect{subsec:HardSpheresPhysics}).

The  model describes $N$ disks of radius $\sigma$ in a rectangular box  with
periodic boundary conditions.
For simplicity, we assume the box to be a square of side length $L$.
The center position of disk $k$ is given by
$\xvec_k = (x_k, y_k)$ and in a \quot{legal} configuration,  any two disks
cannot overlap (get closer than $2
\sigma$), periodic boundary conditions being accounted for.
The sample space $\Omega$ is
now continuous, and the statistical weight of a configuration $\Xvec =
\SET{\xvec_1 \TO \xvec_N}$  is given by
\begin{equation}
 \pi(\Xvec) = \begin{cases}
               1 & \text{if $\Xvec$ is legal} \\
               0 & \text{else}
              \end{cases},
\end{equation}
where, for simplicity, we have omitted the Cartesian $2N$-dimensional measure.
The control parameter of this model is the density $\eta = N \pi \sigma^2 /
L^2$, the fraction of occupied space to the volume of the box.

We consider the global Metropolis algorithm: At each
time step, and starting from a configuration $\Xvec(t) = \SET{\xvec_1 \TO
\xvec_N}$, one random disk $k$ among the $N$ disks  is sampled. A move of disk
$k$ from $\xvec_k$ to a random position inside
the simulation box
$\xvec_k' = \glc \ranb{0, L}
\ranb{0,L} \grc $ is attempted.
If the configuration $\Xvec'$, in which $\xvec$ is
replaced by $\xvec'$ is legal, the move is accepted, and otherwise rejected:
\begin{equation}
 \Xvec(t+1) =
 \begin{cases}
  \SET{\xvec_1 \TO \xvec_{k-1}, \xvec'_k, \xvec_{k+1} \TO \xvec_N}& \text{if
legal}\\
\Xvec(t) & \text{otherwise}
 \end{cases}.
 \label{equ:LabelledDisplacement}
\end{equation}
Here, the new position is chosen within a
square-shaped periodic window of length $L$ around the current position
whereas, in the local Metropolis algorithm, the window size usually has a
length on the scale of the inter-particle distance~\cite{SMAC}.

The \standard coupling for the global Metropolis algorithm uses the following
random element:
\begin{equation}
 \Upsilonvec_t =
\SET{k, \xvec' = \SET{x,y}} =
 \SET{\underbrace{\nran{1}{N}}_{\text{particle index $k$}},
\underbrace{\SET{\ranb{0, L}, \ranb{0, L}}
}_{\text{proposed position $\xvec' = \xvec_k (t+1)$}}}. 
\label{equ:random_element_hs}
\end{equation}
This coupling has been considerably
refined~\cite{Helmuth2022,hayes2014lowerboundscriticaldensity}.

\subsection{Rigorous results for the thermodynamics of hard spheres}
\label{subsec:HardSpheresRigorous}
Rigorous results on hard-disk (and hard-sphere) models are very few. It is
known that the close-packing density  $\eta = \pi/(2\sqrt{3})$ in two
dimensions is characterized by
the hexagonal packing~\cite{Fejes1940}. It thus corresponds to an
essentially unique configuration which has long-range orientational and
positional order. For densities below the close-packing density, the absence of
long-range positional order was established rigorously~\cite{Richthammer2016},
so that there is no crystal (with long-range orientational and positional
order) below close packing.
Indications for a phase transition were first found in the
1960s~\cite{Alder1962,Li2022}. The existence of two phase transitions, and of
three phases (liquid, hexatic, and solid) as a function of density is now well
accepted~\cite{Bernard2011,Li2022}. As in the the \EA
(where the temperature $T$ replaces the inverse density $\eta$ as a control
parameter), a rigorous proof of a transition away from close packing is still
lacking.
At low finite densities, the convergence of the virial expansion was proven
early on~\cite{LebowitzPenrose1964}, establishing the existence of the liquid
phase. It extends up to a density $\eta = 0.70$, and is followed
by a window of coexisting liquid and hexatic regions (see
\tab{tab:hard-disk} below).

\subsection{From path coupling arguments to scaling plots.}
\label{subsec:HardSpheresPathScaling}

\begin{figure}[]
\centering
\includegraphics[align=t,
width=0.49\textwidth]{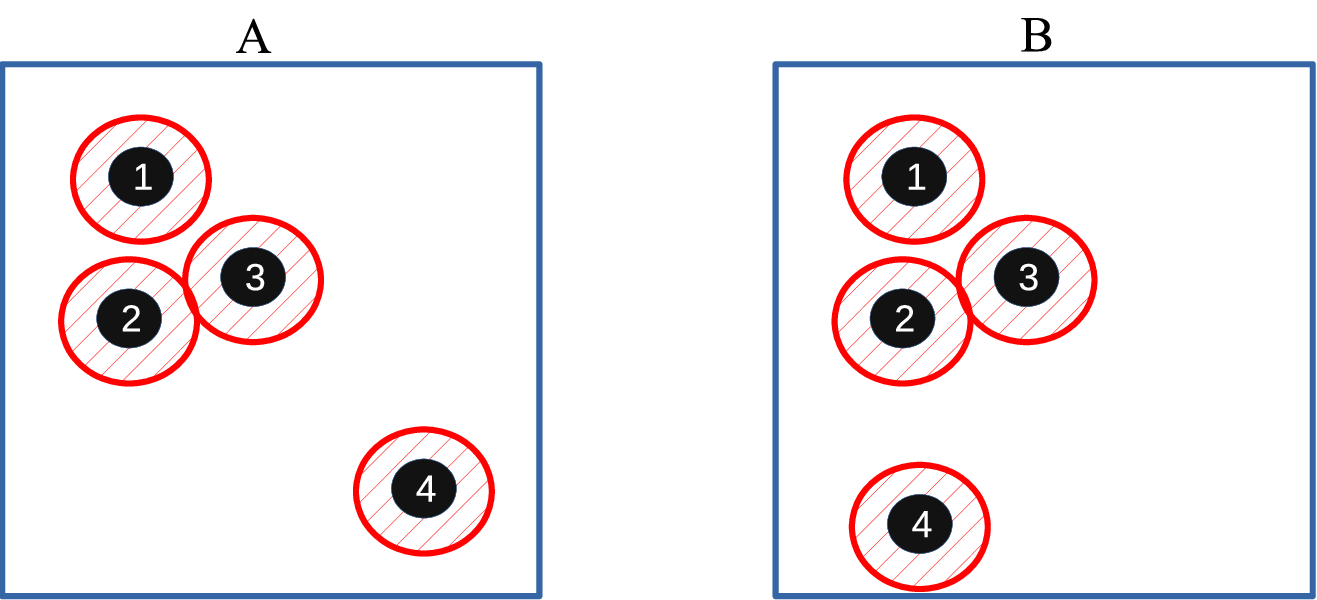}
\caption{ Hard-disk configurations, differing only in disk $j=4$. Under the
\standard coupling, the difference disappears if disk $j$ is moved to a position
outside the \quot{halo} of other disks (see \eq{equ:OneZeroDisks}). It is
increased to two if the move of disk $k \ne j$ would overlap with $j$ in only
one of the configurations (see \eq{equ:OneTwoDisks}). Disks of radius $\sigma$
are shown with their $2\sigma$ halos.}
\label{fig:HardDiskPathCoupling}
\end{figure}

We now consider path coupling for hard disk, using the random map based on
\eq{equ:random_element_hs} and a Hamming metric that counts the number
of different disk positions in any two configurations.
Let $\Xvec^A$ and $\Xvec^B$ be two neighboring hard-disk configurations that
differ in the position of disk $j$ only (see \fig{fig:HardDiskPathCoupling}).
Simplifying a coupling from \REF{Kannanrapidmixing2003}, we use as
the common random element $\SET{k, \Upsilonvec}$ the disk $k$ to be
updated and its new position, both identical for $\Xvec^A$ and $\Xvec^B$.
With probability $1/N$, the disk $j$ is moved (that is, $k=j$). The
move is accepted in both configurations if it stays away (by $2 \sigma$) from
the \quot{halo} of all
remaining disks in both configurations. This yields for the probability
to decrease the Hamming distance from one to zero:
\begin{equation}
 p_{1 \to 0 }  \geq \frac{1}{N} \glb 1 - \frac{N-1}{N} 4 \eta \grb. 
 \label{equ:OneZeroDisks}
\end{equation}
On the other hand, the Hamming distance can be increased from one to two if a
disk different from $j$ is moved less than $2 \sigma$ away (that is, into the
halo), of disk $j$ in one configuration but not in the other. The
probability to increase the Hamming distance from one to two can thus be
bounded as:
\begin{equation}
 p_{1 \to 2 }  \leq \frac{N-1}{N} \glb  \frac{8}{N} \eta \grb. 
 \label{equ:OneTwoDisks}
\end{equation}

Again, for $p_{1 \to 0} > p_{1 \to 2}$, the expected Hamming distance
between $A$ and $B$ decreases after one step, for any two neighboring disk
configurations, which can be assured for
\begin{equation}
 \eta <  \etapath = \frac{1}{12}.
\label{equ:OneTwelve}
\end{equation}
It follows~\cite{BubleyDyer1997} that the Hamming distance between
configurations $A$ and $B$
decreases in expectation at each step two configurations that differ in the
position of only one disk will have after one step, an expected difference which
is smaller than one if the density is smaller than $1/12$.

\begin{figure}
 \centering
\includegraphics[width=0.495\linewidth]{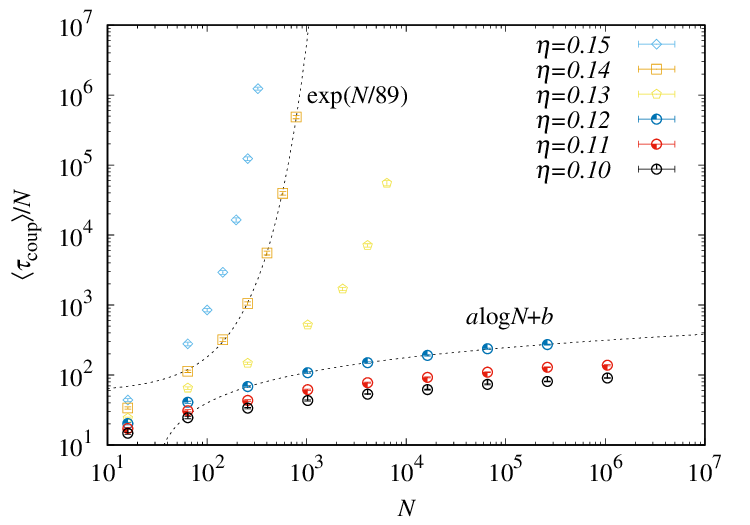}
\includegraphics[width=0.495\linewidth]{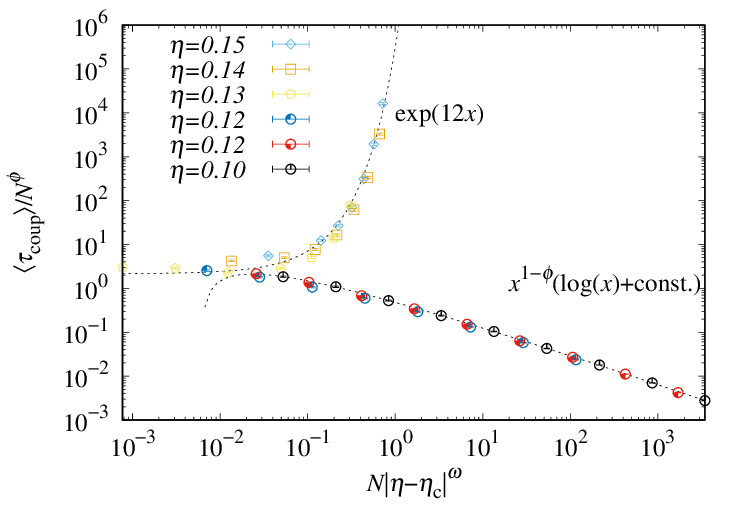}
\caption{\subcap{Left} System-size dependence of the mean coupling
time at various densities in two-dimensional hard disks (data from
\REF{BernardChanalKrauth2010}).
\subcap{Right} Finite-size scaling plot of the coupling time with parameters
$\omega
\simeq 1.6$, $\phi \simeq 1.75$ and $\eta_\mathrm{c} \simeq 0.128$. The two
dotted lines represent the expected high- and low-density asymptotic forms of
the scaling function. }
\label{fig:Ctime_2dHS}
\end{figure}

As with the \EA, we now analyze the mean coupling time of the two-dimensional
hard-disk model under the global Metropolis algorithm with the \standard
coupling of \eq{equ:random_element_hs}. In this case, we reanalyze the data
obtained in \REF{BernardChanalKrauth2010}, which we replot on the lhs of
\fig{fig:Ctime_2dHS}. The analogous scaling ansatz again provides an excellent
fit of the data. The critical exponents do not differ significantly from those
found in the \EA, suggesting the possibility of some underlying universality.
However, uncovering the intricate physical picture behind this similarity
remains an open question for future research. It should be noted that these
critical exponents are not directly related to the critical phenomena of
physical systems in the conventional sense. Rather, they characterize the
\quot{phase transition} in computational algorithms associated with the coupling
of Markov chains. From an algorithmic perspective, these exponents are of
significant interest as they provide insight into the inherent challenges in
achieving fast coupling.

\subsection{Advanced hard-disk couplings, physical implications}
\label{subsec:HardSpheresPhysics}

The coupling approach to the hard-disk system has been intensely studied in
recent years, and the \standard coupling of \eq{equ:random_element_hs} only
provides the simplest possible choice. A number of refined couplings have been
proposed. The one proposed in \REF{Kannanrapidmixing2003} moves disks
differently for the configuration $\Xvec^A$ and $\Xvec^B$, and reaches a
path-coupling density of $1/8$ (see \tab{tab:hard-disk} for an overview).
Building on this coupling, optimizing the metric reaches a limiting density of
$0.154$, which was later improved, for a different algorithm, to $1/6$. Besides
these rigorous bounds, numerical evidence for the birth--death
algorithm~\cite{Wilson2000} points to a coupling density of $\sim
0.3$~\cite{BernardChanalKrauth2010}. These densities, and especially the
rigorously proven ones, are still quite far from the \quot{empirical} transition
density $\eta \sim 0.70$ of the liquid phase, which was only in recent years
understood to be towards a hexatic, and which  bounds on a region $\eta \in
[0.7, 0.76]$ without a homogeneous solution, and then giving rise to a mixture
of the hexatic and the liquid.

\begin{table}[]
    \centering
    \begin{tabular}{c|cl}
        quantity $d$ &  density  & comment \\ \hline
        $\eta_{\text{LP}}$  & $0.03619$ & convergence of
virial expansion, historic first~\cite{LebowitzPenrose1964} \\
        $\etapath$  & $1/12= 0.083$ & naive path-coupling density
        (\eq{equ:OneTwelve}) \\
        $...$  & $1/8 = 0.125$ &
        improved path-coupling~\cite{Kannanrapidmixing2003}\\
        $...$  & $0.154$ & path coupling, optimized
metric~\cite{hayes2014lowerboundscriticaldensity}\\
        $\dots$  & $1/6 = 0.166$ & improved MCMC coupling of
 \REF{Helmuth2022}\\
        $\etacoup$  & $0.128$ & empirical coupling density
        (\fig{fig:Ctime_2dHS}) \\
        $\dots$  & $0.29  $ & empirical birth--death
coupling density (\cite{Wilson2000,BernardChanalKrauth2010} \\
$\eta_{\text{liquid--hex}}$& $0.70 - 0.716$& liquid--hexatic
coexistence~\cite{Alder1962, Bernard2011}\\
$\eta_{\text{hex-solid}}$& $0.72$& hexatic--solid phase transition
~\cite{Bernard2011} \\
$\eta_{\text{pack}}$&  $ \pi / (2 \sqrt{3}) = 0.907$& close-packing crystal
    \end{tabular}
     \caption{Densities in the hard-disk system (see \REF[Eq. (1)]{Li2022}
     for common definitions of densities).
The homogeneous liquid phase empirically extends up a density of $0.70$. the
homogeneous hexatic phase is from $0.716$ to $0.72$. The density range from
$0.70$ to $0.716$ corresponds to phase separation.}
\label{tab:hard-disk}
\end{table}

The crucial connection between fast coupling (thus, fast mixing) and physical
ordering was made, for the hard-sphere case, in \REF{Helmuth2022}, where it was
proven that \bigOb{N \loga{N}} random steps of the global Metropolis algorithm
are insufficient to construct configurations with any kind of long-range order.
Fast
mixing of a single-particle algorithm, even a non-local one, thus implies that
the resulting configuration (which is practically in equilibrium) has
exponential spatial correlation functions. This, to all intents and purposes,
shows the extension of the liquid phase. We believe that it does however not
prove the convergence of the virial expansion~\cite{LebowitzPenrose1964},
because of the possibility of a liquid--liquid phase transition, which cannot be
captured in a mixing-time argument.

\section{Conclusion}
\label{sec:Conclusions}

In this paper, we have discussed computational aspects of two of the most
challenging  models in statistical physics, namely the \EA, and the hard-disk
model. In both these models, there are almost no rigorous results about the
phase transitions in non-trivial physical dimensions, that is, above two
dimensions for the spin model, and above one dimension (away from close packing)
for the particle system. Further connections are that the computational
algorithms
are mostly derivatives of the local-move heat-bath or Metropolis algorithm in
both cases. Cluster algorithms have been developed for both
systems~\cite{Houdayer2001,Dress1995}, but they have not really been useful in
the physically interesting dimensions. Finally, the two models are united by the
fact that they are truly challenging in their physical interpretation: For the
\EA, for a long time, even empirically, there was only a very rough agreed-on
value of the transition temperature from the high-temperature paramagnetic
phase, which was considerably sharpened in recent times only (see
\tab{tab:spin-glass}). No agreement has been reached on the nature of the
low-temperature phase. For the hard-disk model, the now agreed-on transition
scenario~\cite{Bernard2011} was proposed only a decade ago, after more than 50
years of intense
simulation. In that model, even the simplest algorithm, the local Metropolis
algorithm, faces extreme challenges, as its irreducibility and ergodicity cannot
be guaranteed in the constant-volume ensemble~\cite{Boroczky1964,Hoellmer2022}.

In this context, the coupling approach provides an interesting yet incomplete
view on the high-temperature/low-density phases. In the \EA, one can easily
establish the existence of a path-coupling temperature (see \eq{eqn:Tpath}), 
which we
think provides a rigorous lower bound for the extension of the paramagnetic
phase. For
the hard-disk model, the program has been followed through completely, and the
coupling result is the currently best bound for the extension of the liquid
phase. It is fascinating how a result on the speed of a Monte Carlo algorithm
can be derived from the behavior of two Markov chains (that is, from coupling)
and can then be turned into a statement of phase behavior. This fascination was
sensed early on, in the literature on damage spreading that, as we discussed,
naturally connects to the path-coupling approach.

Damage spreading has created an extensive literature in physics but, as we
pointed out, has concentrated on the specific random-share protocol which,
translated to the hard-disk context, gives the very low bounding density of
\eq{equ:OneTwelve}. In particle systems, there has been much progress from
improved couplings and optimized metrics (see \tab{tab:hard-disk}), which we
hope can be ported to spin glasses and, more generally, to disordered systems.
It would be interesting to see whether our scaling approach can be applied to
these more advanced couplings.

\section*{Conflict of Interest Statement}

The authors declare that the research was conducted in the absence of any
commercial or financial relationships that could be construed as a potential
conflict of interest.

\section*{Author Contributions}
KH and WK conducted the literature review and performed numerical simulations.
Both authors contributed to the writing of the manuscript.

\section*{Funding}
This research was supported by a grant from the Simons Foundation
(Grant 839534, MET).
This work was also supported by JSPS KAKENHI Grant Nos. 23H01095, 
and JST Grant Number JPMJPF2221.
This research was conducted within
the context of the International Research Project
\quot{\emph{Non-Reversible Markov chains, Implementations and
Applications}}.

\section*{Acknowledgments}
We thank J. L. Lebowitz for an inspiring discussion.
KH would like to thank the Ecole Normale Sup\'{e}rieure ENS for their kind hospitality
for a research stay, which provided a productive environment and variable support for
completion of this work.


\bibliographystyle{Frontiers-Harvard} 
\bibliography{General,DamageMix}

\end{document}